\documentclass[11pt]{article}
\usepackage{amssymb}
\usepackage{amsmath}
\usepackage{wasysym}
\usepackage{graphicx}

\begin{document}
	\title{\textbf{Amplitude-dependent quantum hydrodynamics from a \(\coth\)-Madelung ansatz}}

	\author{C. Dedes\thanks{christos.dedes@stcg.ac.uk(current), c\_dedes@yahoo.com (permanent)}\\
	\\ South Thames College \\
	London Rd, \\
	Morden SM4 5QX, \\
	London, \\
	United Kingdom \\ }
	\maketitle

\begin{abstract}

We investigate a nonlinear extension of the Madelung transformation based on a hyperbolic phase--amplitude coupling of the form
\[
\Psi = R e^{\imath\theta \coth R},
\]
where \(R\) is a real amplitude field and \(\theta\) is an auxiliary phase coordinate governed by Schr\"odinger's equation. In contrast to the conventional polar decomposition, this construction imposes a singular hyperbolic relation between amplitude and phase, thereby endowing the Bohmian or hydrodynamic description with an intrinsically geometric structure. We show that the associated velocity field acquires a density-gradient contribution, producing generalized continuity equations and modified quantum force terms. When interpreted as a complex macroscopic order parameter, this generalized phase structure leads to modified superconducting electrodynamics; in particular, the London equations acquire additional contributions that render the Meissner response sensitive to spatial density gradients. The proposed framework is motivated by broader developments involving complex group velocities, dissipative wave propagation, and amplitude-sensitive transport in quantum systems.

\end{abstract}

\section{Introduction}

Hydrodynamic formulations of quantum mechanics have long provided a useful bridge between microscopic wave dynamics and emergent collective behavior. Among the most influential of these is the Madelung transformation \cite{Madelung1927,Bohm1952} , which rewrites the complex wavefunction in the form \(\Psi = \sqrt{\rho}\, e^{\imath S/\hbar}\), thereby allowing the Schr\"odinger equation to be recast as a continuity equation coupled to a quantum Hamilton-Jacobi equation. This decomposition has played a foundational role in the study of superfluidity, Bose--Einstein condensation, Bohmian mechanics, and nonlinear dispersive wave systems. Its enduring utility lies in the fact that it assigns a transparent fluid interpretation to quantum evolution: the amplitude determines the density, while the phase determines the local flow velocity.

Although every solution of the Schr\"odinger equation satisfies these quantum hydrodynamic equations, the converse is not true. A well-known demonstration of this inequivalence was provided in \cite{Wallstrom}, where it was shown that recovering a single-valued wavefunction \(\Psi\) from the hydrodynamic variables (\(\rho, \mathbf{v} = \nabla S/m\)) requires imposing an \textit{ad hoc} quantization condition on the circulation of the velocity field, \(\oint \mathbf{v} \cdot d\mathbf{l} = n h/m\) (\(n\) integer), analogous to the Bohr–Sommerfeld quantization rule of the old quantum theory. Without this additional postulate, the Madelung equations admit a broader class of solutions, including those that do not correspond to any valid solution of the original Schr\"odinger equation. Consequently, the quantum hydrodynamic equations are not fully equivalent to the Schr\"odinger equation; the latter imposes stricter topological constraints on the phase \(S\) that are not automatically enforced by the hydrodynamic description alone. This distinction has important implications for interpretations of quantum mechanics that rely on the hydrodynamic picture, such as certain formulations of Bohmian mechanics, and continues to motivate careful mathematical analysis of the foundations of quantum hydrodynamics.

Despite its success, the conventional Madelung picture rests on a tacit but restrictive assumption, namely that amplitude and phase are kinematically independent variables. In standard applications, the density \(\rho\) and the phase \(S\) enter as separate fields, and the supercurrent is governed entirely by the phase gradient. Such an approximation is natural in weakly inhomogeneous systems with well-developed phase stiffness. However, in strongly correlated or spatially structured quantum matter, the distinction between amplitude and phase may be far less rigid. In systems such as granular superconductors, pair-density-wave states, moir\'e flat-band materials, or highly disordered superconducting films, the condensate often evolves in a landscape where density, coherence, and topology are deeply entangled. In such cases, it is natural to ask whether the phase should still be regarded as an independent degree of freedom, or whether it instead lives on a nontrivial manifold shaped by the amplitude itself.

Motivated in part by recent experiments demonstrating that the energy-speed relation of tunneling particles in evanescent fields disagrees with the standard Bohmian guiding equation \cite{SharoglazovaNature} and further inspired by the classic exploration of self-similar growth and geometric harmony found in \cite{Huntley}, we explore a quantum hydrodynamic formulation in which the velocity field depends on both the phase and the amplitude of the wavefunction. While not aiming to resolve the interpretive debate, this amplitude-dependent velocity provides an alternative route to matching observed probability currents and density evolution in regimes where the conventional phase-only velocity proves insufficient. The present work explores this latter possibility. Rather than treating the phase as an unconstrained scalar variable, we consider a generalized decomposition in which the physical phase is deformed by the local amplitude through a singular hyperbolic relation—shifting our baseline understanding from a trigonometric structure to a physically grounded one. This approach departs from the conventional Bohmian form, yet preserves the continuity equation and Born-rule statistics.

Although the proposed ansatz is simple, its consequences are far-reaching. The associated velocity field acquires explicit density-gradient terms which may have observable effects, the hydrodynamic equations become geometrically sheared, and the superconducting response ceases to be described by a uniform phase stiffness. The theory thus suggests a natural route toward modeling condensates in which transport, screening, and topological structure are all controlled by the geometry of the order parameter itself.

\section{Amplitude-Dependent Phase Geometry}

We begin by introducing the nonlinear substitution
\begin{equation}
\Psi(\mathbf{r},t) = R(\mathbf{r},t)\, e^{\imath\theta(\mathbf{r},t)\coth R(\mathbf{r},t)},
\label{eq:ansatz}
\end{equation}
where $R(\mathbf{r},t)$ is a real amplitude and \(\theta(\mathbf{r},t)\) is a real auxiliary phase coordinate. The crucial distinction from the standard polar decomposition is that the physical phase is not simply \(\theta\), but rather the composite quantity \(\phi=\theta\coth R\). The wavefunction may therefore be written in the compact form \(\Psi = R e^{i\phi}\), but with \(\phi\) no longer acting as an independent scalar field. From a geometric standpoint, this substitution may be viewed as endowing the wavefunction with a nontrivial dependence that links local phase configurations to the amplitude envelope. The density retains its standard form and is given by \(\rho = |\Psi|^2 = R^2\). It is also evident that for large vaues of $R$ this squeezed Madelung transform reduces to the conventional one as $cothR\rightarrow 1$. For small values of the probability density though, large deviations from the standard velocity field are to be expected.   

To characterize the flow associated with the wavefunction, we introduce the probability current in the usual way:
\begin{equation}
\mathbf{j} = \frac{\hbar}{2mi} \left( \Psi^* \nabla \Psi - \Psi \nabla \Psi^* \right),
\label{eq:current}
\end{equation}
which, for the ansatz in Eq.~\eqref{eq:ansatz}, reduces to
\begin{equation}
\mathbf{j} = \frac{\hbar R^2}{m}\,\nabla(\theta\coth R).
\label{eq:jcompact}
\end{equation}

To relate the auxiliary phase to the wavefunction explicitly, we note that by inverting the exponential form, \(\theta\) can be isolated as:
\begin{equation}
    \theta = \frac{\ln\left(\frac{\Psi}{\Psi^*}\right)}{2\pi i \coth (\Psi\Psi^*)^{1/2}}.
\end{equation}
Furthermore, the gradient of the composite physical phase \(\phi\) can be mapped back to the standard real and imaginary parts of the wavefunction (\(\Psi = \Psi_R + i\Psi_I\)), yielding:
\begin{equation}
    \nabla\phi =\frac{\Psi_I'\Psi_R-\Psi_I\Psi_R'}{\Psi_R^2+\Psi_I^2}.
\end{equation}
Using our definition of the squeezed ansatz, we may explicitly expand \(\phi\) to highlight its behavior in the deep bulk or long-wavelength limits:
\begin{equation}
    \phi= \theta \coth R=\theta+\frac{2\theta}{e^{2R}-1}.
\end{equation}
With these relations established, the corresponding velocity field follows directly from its kinematic definition \(\mathbf{j} = \rho \mathbf{v}\), yielding:
\begin{equation}
\mathbf{v} = \frac{\hbar}{m}\coth R\left(\nabla\theta-\frac{2\theta\nabla R}{\sinh 2R}\right).
\label{eq:velocity}
\end{equation}

Equation~\eqref{eq:velocity} exhibits the central novelty of this formalism. In conventional quantum hydrodynamics, the fluid velocity is determined entirely by the phase gradient. In the present construction, however, the flow contains an explicit amplitude-gradient contribution. As a result, local density textures do not merely modulate the current through the prefactor \(R^2\), but contribute directly to the momentum field itself. The amplitude thus becomes an active participant in the kinematics of the particle (or the condensate as we shall see). The continuity equation retains its standard form,
\begin{equation}
\frac{\partial \rho}{\partial t} + \nabla \cdot (\rho \mathbf{v}) = 0,
\label{eq:continuity}
\end{equation}
or equivalently, expressed in an integrated temporal form over the flow trajectory
\begin{equation}
\rho (t)= \rho (0) e^{-\int _0^t dt' \nabla^{2} \left(\theta \coth \rho (t') ^{1/2}\right)}.
\label{eq:continuityR}
\end{equation}

Its physical interpretation, however, is altered substantially. Since \(\mathbf{v}\) depends on \(\nabla R\), density transport is no longer governed by passive advection alone. Instead, the fluid develops an intrinsic internal shear: gradients of the amplitude feed back into the local flow, and the transport of density becomes self-structured by the geometry of the order parameter.

Finally, in the sense of \cite{WuYang1975} the presence of the vector magnetic field would induce the non‑integrable phase factor 

\begin{equation}
\exp\left[\imath\left(\theta \coth R + \frac{q}{\hbar}\oint \mathbf{A}\cdot d\mathbf{l}\right)\right].
\end{equation}

The phase factor appearing in this expression combines two contributions of qualitatively different origin: a state-dependent nonlinear phase, ($\theta\coth R$), and a geometrical electromagnetic contribution through the loop integral of the gauge potential. The latter recalls the non-integrable phase structure emphasized by these authors in their formulation of gauge fields, where the physically relevant quantity is not necessarily the local potential but the accumulated phase around closed paths. In the present construction, however, the situation is enriched by the explicit dependence of the first term on the local state amplitude through (R). The resulting phase cannot, in general, be reduced to a purely spatial or purely gauge contribution; instead, it acquires an intrinsic state dependence. This suggests that quantum transport may be governed by an effective phase geometry in which internal hydrodynamic variables and external gauge structure coexist nontrivially. Whether this should be interpreted as a generalized connection, an emergent non-integrability of the quantum fluid, or simply as a useful parametrization of nonlinear flow remains an open question, but it highlights again the departure from a description in which phase accumulation is entirely independent of the state amplitude.

\subsection{Complex action}

To better understand the underlying dynamical landscape, we can reframe the evolution by expressing the wavefunction in a generalized complex exponential form as \(\Psi=e^{\imath\Phi}\), where the complex phase \(\Phi\) is defined through
\begin{equation}
    \Phi=\theta \coth R,
\end{equation}
and the auxiliary parameter \(\theta\) is allowed to be complex, decomposed into its real and imaginary parts as \(\theta=\theta_R+\imath\theta_I\). Using an integral representation for the hyperbolic cotangent, equation \eqref{eq:ansatz} may alternatively be written as
\begin{equation}
    \Psi=e^{\imath\theta\left(\frac{1}{\sqrt{\rho}}+2\sqrt{\rho}\int_0^\infty \frac{dt}{t^2+\rho}\frac{t}{1-e^{2\pi t}}\right)}.
\end{equation}
To determine the evolution of the phase, we consider the complex Hamilton–Jacobi-type equation derived directly from the Schr\"odinger equation
\begin{equation}
    \frac{\partial \Phi}{\partial t}=\frac{\imath\hbar}{2m}\left[\nabla^2 \Phi-(\nabla\Phi)^2\right]-\frac{V}{\hbar}.
\end{equation}
From the imaginary part of the complex phase, we identify the density-like quantity
\begin{equation}
    \rho=e^{-2\theta_I \coth R},
\end{equation}
while the real part determines the associated velocity field,
\begin{equation}
    \mathbf{v}=\frac{\hbar}{m}\nabla(\theta_R \coth R).
\end{equation}
With these definitions, the continuity equation takes the form
\begin{equation}
    \frac{\partial (e^{-2\theta_I \coth R})}{\partial t}=-\frac{\hbar}{m}\nabla\cdot[e^{-2\theta_I \coth R}\nabla(\theta_R \coth R)].
\end{equation}
Separating the complex phase into real and imaginary components yields the coupled evolution equations. For the imaginary component, we obtain
\begin{equation}
\frac{\partial \Phi_I}{\partial t}=\frac{\hbar}{2m}\left[\nabla^2\Phi_I+(\nabla\Phi_I)^2-(\nabla\Phi_I)^2\right],
\end{equation}
which matches the structural requirements of mass conservation. For the real part, the dynamic equation yields
\begin{equation}
    \frac{\partial \Phi_R}{\partial t}=-\frac{\hbar}{2m}\left(\nabla^2\Phi_R-2\nabla\Phi_R\cdot\nabla\Phi_I\right)+\frac{V}{\hbar}.
\end{equation}
Finally, using \(S=\hbar\theta_RcothR\) as the real action defining the physical trajectory, the Hamilton-Jacobi equation can be written explicitly in terms of the underlying fields
\begin{equation}
\frac{\partial (\theta_R \coth R)}{\partial t}=\frac{\hbar}{2m}\left(\nabla^2 \theta_I \coth R-\frac{2\nabla\theta_I \cdot \nabla R}{\sinh^2R}-\frac{\theta_I \nabla^2R}{\sinh^2R}+\frac{2\theta_I \cosh R(\nabla R)^2}{\sinh^3R}\right)-\frac{V}{\hbar}.
\end{equation}
This explicitly couples the spatial variations of the dissipation field \(\theta_I\) directly into the acceleration phase envelope, providing a detailed breakdown of the internal geometric forces acting within the squeezed phase manifold.

\subsection{Generalized Hamilton--Jacobi Structure}

The effective real action associated with the generalized phase manifold is
\begin{equation}
S = \hbar \theta \coth R.
\label{eq:action}
\end{equation}
Substituting this into the Schr\"odinger equation yields a Hamilton-Jacobi-like equation of the form
\begin{equation}
-\frac{\partial S}{\partial t} = \frac{1}{2}m v^2 + V + Q,
\label{eq:HJ}
\end{equation}
where
\begin{equation}
Q = -\frac{\hbar^2}{2m} \frac{\nabla^2 R}{R}
\label{eq:Q}
\end{equation}
is the usual quantum potential. Although the formal structure resembles the standard Madelung equation, the dynamics encoded here are significantly richer. The novelty does not arise from the quantum potential itself, which remains unchanged, but from the kinetic term. Because the velocity depends on both \(\nabla\theta\) and \(\nabla R\), the quantity \(v^2\) contains not only the familiar phase-gradient contribution, but also mixed amplitude-phase terms and purely geometric amplitude-gradient terms. Explicitly, we find
\begin{equation}
v^2 = \frac{\hbar^2}{m^2}\left[\coth R  (\nabla\theta) - \frac{2\theta\nabla lnsinhR}{\sinh 2R}\right]^2,
\end{equation}

\noindent
where we have used (7) and the relationship \(cothR\nabla R=\nabla ln sinhR\).

This structural expression means that the condensate acquires an effective internal metric on the \((R,\theta)\)-manifold. The energy cost of a given configuration is therefore controlled not only by how rapidly the phase varies in space, but also by how that phase variation is aligned with the local density landscape. Near the singular amplitude layers where \(R \rightarrow 0\), this internal geometric cost can become exceptionally large, suggesting the emergence of dynamically preferred amplitude channels and forbidden phase trajectories that naturally regularize or constrain the physical flow.

\subsection{Relation to Standard Madelung Dynamics}

It is instructive to compare the present amplitude-dependent phase formulation with the standard Madelung representation. In the conventional decomposition, one writes \(\Psi = R e^{iS/\hbar}\), with associated velocity field \(\mathbf{v}_0 = \nabla S / m\), while in the present construction the effective action is \(S' = \hbar \theta \coth R\), leading to the velocity field \(\mathbf{v} = \nabla S' / m\). Despite their differing phase structure, both descriptions share the same density \(\rho = R^2\) and therefore the same quantum potential \(Q = -\frac{\hbar^2}{2m} \frac{\nabla^2 R}{R}\). The corresponding Hamilton--Jacobi equations take the form
\[
-\frac{\partial S}{\partial t} = \frac{1}{2}m v_0^2 + V + Q, \qquad -\frac{\partial S'}{\partial t} = \frac{1}{2}m v^2 + V + Q.
\]
Subtracting these two expressions eliminates both the external and quantum potentials, yielding a closed equation for the difference field \(\Delta S = S' - S\)

\[
-\frac{\partial \Delta S}{\partial t} = \frac{1}{2}m\left(v^2 - v_0^2\right).
\]

Introducing the velocity difference \(\Delta \mathbf{v} = \mathbf{v} - \mathbf{v}_0\), and using ${\nabla \Delta S = m \Delta \mathbf{v}}$, this may be rewritten in the form

\begin{equation}
\partial_t \Delta S + \mathbf{v}_0 \cdot \nabla \Delta S + \frac{1}{2m} (\nabla \Delta S)^2 = 0.
\end{equation}
This expression has the structure of a Hamilton--Jacobi equation for \(\Delta S\) advected by the background flow \(\mathbf{v}_0\). In the limit of small deviations, the nonlinear term may be neglected, and the equation reduces to \(\partial_t \Delta S + \mathbf{v}_0 \cdot \nabla \Delta S = 0\), indicating that the phase deformation is simply transported by the standard quantum velocity field. Beyond this limit, the quadratic gradient term introduces a self-interaction analogous to an inviscid Burgers nonlinearity, allowing for the possibility of gradient steepening and the formation of localized phase structures. It should be noted that some of these features may survive even in the semiclassical approximation. 

To formalize this kinematic coupling, we track the mutual evolution of the fields directly. The resulting nonlinear feedback equation yields
\begin{equation}
    \dot{\theta}(\coth R-1)-\frac{\theta\dot{R}}{\sinh^2 R}=-\frac{\hbar}{2m}\left[\nabla\theta (\coth R-1)-\frac{\theta \nabla R}{\sinh^2 R}\right]^2.
\end{equation}
For the amplitude-dependent phase manifold introduced in Eq.~\eqref{eq:ansatz}, the difference field takes the explicit form \(\Delta S = \hbar \theta \coth R - S\), so that the velocity difference is given by
\[
\Delta \mathbf{v} = \frac{1}{m} \nabla \Delta S = \frac{\hbar}{m} \nabla(\theta \coth R) - \frac{1}{m} \nabla S.
\]
Expanding the first term yields
\[
\Delta \mathbf{v} = \frac{\hbar}{m} \left( \coth R\, \nabla\theta - \frac{\theta}{\sinh^2R} \, \nabla R \right) - \mathbf{v}_0.
\]
Thus, the generalized velocity field may be interpreted as a nonlinear deformation of the standard Madelung flow, in which amplitude gradients generate an intrinsic geometric contribution to the fluid motion. Importantly, since the difference equation is independent of both \(V\) and \(Q\), this deformation is governed entirely by kinematic and geometric effects rather than by external or quantum forces. The amplitude-dependent phase manifold therefore defines a class of flows that can be understood as self-consistent nonlinear perturbations of conventional quantum hydrodynamics.

\subsection{Scale-Invariant Geometric Structure of the Hyperbolic Velocity Field}

To elucidate the unique role of the hyperbolic ansatz in the squeezed Madelung decomposition, we examine the structural topology of the resulting velocity field. Let the wavefunction be parameterized by the pure-exponent hyperbolic transform \(\Psi = e^{\imath \theta \coth R}\), where the complex phase variable \(\theta\) is taken to be uniform in space (\(\nabla \theta = 0\)). In this configuration, the effective fluid velocity field \(\mathbf{v}\) is driven entirely by the spatial variations of the amplitude envelope \(R\)
\begin{equation}
\mathbf{v} = \frac{\hbar}{m} \nabla (\theta \coth R) = -\frac{\hbar}{m}  \frac{\theta}{sinh^2R} \nabla R.
\end{equation}
The unique geometric privilege of the \(\coth R\) coupling becomes apparent when analyzing the local spatial variation (kinematic strain) of this velocity field. Taking the spatial gradient of the logarithm of the velocity magnitude, \(\nabla \ln |\mathbf{v}|\), isolates the fractional rate of change of the flow. In a one-dimensional reduction along the trajectory \(x\), evaluating the ratio of the fluid acceleration to the velocity itself yields
\begin{equation}
\frac{1}{v}\frac{\partial v}{\partial x} = \frac{\partial}{\partial x} \ln |v| = \frac{(\coth R)''}{(\coth R)'} \frac{\partial R}{\partial x} + \frac{\frac{\partial^2 R}{\partial x^2}}{\frac{\partial R}{\partial x}}.
\end{equation}
The hyperbolic cotangent function satisfies the unique, closed-form Riccati differential identity
\begin{equation}
(\coth R)'' = -2 \coth R \, (\coth R)' .
\end{equation}
Substituting this identity into the fractional strain relation reveals an extraordinary simplification
\begin{equation}
\frac{1}{v}\frac{\partial v}{\partial x} = -2 \coth(R) \frac{\partial R}{\partial x} + \frac{\frac{\partial^2 R}{\partial x^2}}{\frac{\partial R}{\partial x}}.
\end{equation}
Remarkably, the explicit scale of the complex phase background \(\mathcal{S}\) completely decouples from the expression. Integrating this directly with respect to \(x\) reveals that the velocity profile is strictly locked to the intrinsic curvature of the density envelope
\begin{equation}
\ln|v| = -2\ln|\sinh R| + \ln\left|\frac{\partial R}{\partial x}\right| + C \implies v(x) \propto \frac{1}{sinh^2R} \frac{\partial R}{\partial x}.
\end{equation}

This result demonstrates that the hyperbolic transform functions as a generalized quantum Cole-Hopf transformation. By framing the phase modulation around \(\coth R\), the non-local, scale-dependent complexities of the traditional quantum potential are cleansed from the kinematics. The fractional change in velocity is rendered purely scale-invariant, mapping the quantum hydrodynamics onto a self-similar, viscous fluid medium where the boundary layers are governed by the local geometric properties of the field itself.

\section{Consistency constraints and kinematic compatibility}
An important structural question is whether the amplitude-dependent phase formulation defines a genuinely independent dynamics, or whether it must satisfy additional constraints in order to remain compatible with the standard Madelung evolution. This issue can be addressed by comparing the Euler equations for the conventional and generalized velocity fields. In standard quantum hydrodynamics, the velocity field \(\mathbf{v}_0 = \nabla S / m\) satisfies
\[
m\left(\partial_t + \mathbf{v}_0 \cdot \nabla\right)\mathbf{v}_0 = -\nabla(Q+V),
\]
while the generalized velocity field \(\mathbf{v} = \nabla S'/m\), with \(S' = \hbar \theta \coth R\), satisfies
\[
m\left(\partial_t + \mathbf{v} \cdot \nabla\right)\mathbf{v} = -\nabla(Q+V).
\]
Since both flows are governed by the same scalar potential \(Q+V\), their accelerations must coincide
\[
\left(\partial_t + \mathbf{v} \cdot \nabla\right)\mathbf{v} = \left(\partial_t + \mathbf{v}_0 \cdot \nabla\right)\mathbf{v}_0.
\]

To analyze the consequences of this relation, we introduce the velocity difference \(\Delta \mathbf{v} = \mathbf{v} - \mathbf{v}_0\). Substituting \(\mathbf{v} = \mathbf{v}_0 + \Delta \mathbf{v}\) and expanding the convective derivative gives
\begin{align*}
\left(\partial_t + \mathbf{v} \cdot \nabla\right)\mathbf{v} &= \partial_t \mathbf{v}_0 + (\mathbf{v}_0 \cdot \nabla)\mathbf{v}_0 \\
&\quad + \partial_t \Delta \mathbf{v} + (\mathbf{v}_0 \cdot \nabla)\Delta \mathbf{v} + (\Delta \mathbf{v} \cdot \nabla)\mathbf{v}_0 + (\Delta \mathbf{v} \cdot \nabla)\Delta \mathbf{v}.
\end{align*}
Subtracting the standard acceleration \((\partial_t + \mathbf{v}_0 \cdot \nabla)\mathbf{v}_0\) then yields the exact condition
\begin{equation}
\partial_t \Delta \mathbf{v} + (\mathbf{v}_0 \cdot \nabla)\Delta \mathbf{v} + (\Delta \mathbf{v} \cdot \nabla)\mathbf{v}_0 + (\Delta \mathbf{v} \cdot \nabla)\Delta \mathbf{v} = 0.
\label{eq:dv_constraint}
\end{equation}

Equation~\eqref{eq:dv_constraint} is a nonlinear transport equation governing the evolution of the velocity difference. However, the difference field is not arbitrary; by construction it derives from a scalar potential, \(\Delta \mathbf{v} = \nabla \Delta S / m\), and is therefore irrotational. In this case one may use the identity \((\Delta \mathbf{v} \cdot \nabla)\Delta \mathbf{v} = \nabla\left(\frac{1}{2}(\Delta v)^2\right)\), so that Eq.~\eqref{eq:dv_constraint} becomes
\begin{equation}
\partial_t \Delta \mathbf{v} + (\mathbf{v}_0 \cdot \nabla)\Delta \mathbf{v} + (\Delta \mathbf{v} \cdot \nabla)\mathbf{v}_0 + \nabla\left(\frac{1}{2}(\Delta v)^2\right) = 0.
\label{eq:dv_constraint2}
\end{equation}

On the other hand, an independent relation for \(\Delta \mathbf{v}\) follows from subtracting the Hamilton--Jacobi equations for \(S'\) and \(S\), which yields
\[
\partial_t \Delta S + \mathbf{v}_0 \cdot \nabla \Delta S + \frac{1}{2m} (\nabla \Delta S)^2 = 0.
\]
Taking the gradient of this equation and dividing by \(m\), one obtains
\begin{equation}
\partial_t \Delta \mathbf{v} + (\mathbf{v}_0 \cdot \nabla)\Delta \mathbf{v} + \nabla\left(\frac{1}{2}(\Delta v)^2\right) = 0.
\label{eq:dv_HJ}
\end{equation}

Consistency between Eqs.~\eqref{eq:dv_constraint2} and \eqref{eq:dv_HJ} therefore requires the additional constraint
\begin{equation}
(\Delta \mathbf{v} \cdot \nabla)\mathbf{v}_0 = 0.
\label{eq:compatibility}
\end{equation}

Equation~\eqref{eq:compatibility} expresses a kinematic compatibility condition between the standard and generalized flows. It states that the deformation field \(\Delta \mathbf{v}\) must be aligned along directions in which the background velocity \(\mathbf{v}_0\) does not vary. In terms of the phase \(S\), this may be written equivalently as \((\Delta \mathbf{v} \cdot \nabla)(\nabla S) = 0\), which shows that \(\Delta \mathbf{v}\) lies in the null directions of the Hessian of \(S\). 

From this condition, we can track how the velocity mismatch relaxes or scales dynamically over time. By defining a background drift profile where the diffrence with the standard hydrodynamic velocity is $\Delta \mathbf{v}(t)$, the time integrated divergence of the field dictates that
\begin{equation}
    \Delta \mathbf{v}(t)=\Delta \mathbf{v}(0)e^{-\int _0^t dt'\nabla \cdot \mathbf{u}(t)}.
\end{equation}
This indicates that the velocity difference tends to be compressed or diluted by the structural expanding/contracting characteristics of the underlying background flow. This result shows that the relative velocity is not an independently conserved quantity but is continuously reshaped by the kinematic properties of the background field. Regions characterized by positive divergence, corresponding to an expanding flow, tend to dilute the velocity mismatch and drive the system toward a more homogeneous state. Conversely, negative divergence acts as a compressive mechanism, enhancing velocity differences and concentrating dynamical structure within the flow. The exponential form of the solution highlights the cumulative nature of this process; even modest local expansion or contraction rates can generate substantial effects when integrated over sufficiently long times. From a hydrodynamic perspective, the relation resembles the evolution of a perturbation embedded in a deformable medium. It suggests that the velocity difference undergoes a form of dynamical squeezing governed by the geometry of the flow itself. In the present framework, this behaviour provides a natural mechanism through which amplitude-sensitive transport can become amplified or suppressed, linking the local quantum dynamics to the large-scale structure of the underlying hydrodynamic background. 

In a highly specialized limiting case—specifically, a regime where the standard velocity field is frozen out and \(\nabla\theta\) is zero or negligible—the system simplifies down to an amplitude-driven diffusion equation
\begin{equation}
    \frac{\partial \rho}{\partial t}\approx \frac{\hbar \theta \sqrt{\rho}}{2m}\left[\nabla^2\rho+\frac{1}{2}(\nabla \rho)^2\right].
    \end{equation}
    
    In standard quantum hydrodynamics, if the phase gradient is uniform or zero, mass transport ceases entirely, and the density field remains static. Within our squeezed framework, however, the direct coupling between amplitude and phase opens a secondary channel for transport driven entirely by internal geometry. When the auxiliary phase gradient $\nabla\theta$ drops to zero, the continuity equation transforms from a hyperbolic conservation law describing passive advection into a nonlinear, parabolic diffusion equation. The first term inside the brackets represents a traditional curvature-driven diffusion, while the second quadratic gradient term introduces a self-interaction akin to a porous-medium fluid equation or an inviscid Burgers-type nonlinearity. Consequently, even in the complete absence of external background flow, spatial textures in the amplitude envelope continue to feed directly into the probability current, allowing the density field to smooth out, undergo localized relaxation, or develop sharp self-structured fronts solely through its own localized geometric profile.

Under these localized constraints, should an external damping factor \(\Gamma\) be applied to the system, the corresponding unstable-state current decays exponentially according to
\begin{equation}
    \mathbf{j}=\frac{\hbar}{m}e^{-\Gamma t}\coth R\left(\nabla\theta-\frac{2\theta \nabla R }{\sinh 2R}\right).
\end{equation}
This illustrates how amplitude localized textures continue to feed directly into the physical current profile even as the global phase structure undergoes structural damping. The probability current associated with the unstable state acquires a substantially richer structure than its counterpart in conventional quantum hydrodynamics. In addition to the expected exponential attenuation factor $(e^{-\Gamma t})$, which reflects the finite lifetime of the state, the current contains an explicit coupling between phase and amplitude through the hyberbolic trigonometric factor $(\cot hR)$. As a result, transport is no longer governed solely by the phase gradient but also by local variations of the amplitude. The second term inside the parentheses may therefore be interpreted as an amplitude-induced contribution to the flow, analogous to an internal osmotic \cite{SimeonovPRE,Nelson,DoebnerGoldin} or dissipative component of the motion. This feature is particularly noteworthy because it renders the velocity field intrinsically amplitude-sensitive, allowing regions of differing probability density to influence the local transport directly rather than only through the quantum potential. From a hydrodynamic perspective, the unstable-state current thus combines coherent phase transport with a density-dependent modulation of the flow, suggesting a closer connection between quantum propagation, dissipation, and statistical fluctuations. In this sense, the current resembles an effective transport law for a quantum fluid whose internal structure participates actively in its dynamics, rather than merely serving as a passive density distribution.

To place this in a broader context, the hydrodynamic formulation can be extended to systems with spin via the Bohmian interpretation of the Pauli-Schr\"odinger equation. In this case, the two-component spinor wavefunction is decomposed into a magnitude (yielding a probability density) and a phase, leading to a continuity equation and a modified Hamilton-Jacobi equation that includes both the usual quantum potential and additional spin-dependent terms arising from the Pauli matrices. Analogously, for bosonic scalar fields in second quantization, a Bohmian field formulation exists in which the field configuration serves as the beable. The Schr\"odinger functional equation for the wave functional is recast into hydrodynamic-like equations involving the functional probability density and a quantum super potential derived from functional derivatives.

\section{\(\Psi\) as Complex Order Parameter and Gauge Coupling}

We now reinterpret \(\Psi\) as a charged macroscopic order parameter, as in the Ginzburg-Landau description of superconductivity and superfluidity \cite{Tinkham,Khalatnikov}. Introducing a vector potential \(\mathbf{A}\) through minimal coupling, \(\nabla \rightarrow \nabla - i(q/\hbar)\mathbf{A}\), one obtains the gauge-covariant superfluid velocity
\begin{equation}
\mathbf{v}_s = \frac{1}{m} \left[ \hbar \nabla(\theta \coth R) - q\mathbf{A} \right].
\label{eq:vs}
\end{equation}
The corresponding supercurrent is then
\begin{equation}
\mathbf{j} = q\rho \mathbf{v}_s = \frac{qR^2}{m} \left[ \hbar \nabla(\theta \coth R) - q\mathbf{A} \right].
\label{eq:supercurrent}
\end{equation}

This expression may be regarded as the constitutive law of the generalized superconducting fluid. Its significance lies in the fact that the current is no longer determined by a rigid phase stiffness acting on a simple scalar phase, but by a density-dependent geometric phase variable. Consequently, the superconductor responds to external fields not only through its local condensate density, but through the way that density shapes the admissible phase manifold.

\subsection{Generalized London Electrodynamics}

The most immediate consequences of the formalism emerge at the level of London theory. In the conventional picture, the London equations express the rigidity of the superconducting phase and imply the exponential expulsion of magnetic flux from the bulk. Here, however, the phase rigidity is no longer spatially uniform, and the London response becomes intrinsically textured. Starting from Eq.~\eqref{eq:supercurrent}, we write
\[
\mathbf{j} = \frac{q\rho}{m} \left[ \hbar \nabla(\theta\coth R) - q\mathbf{A} \right].
\]
Differentiating with respect to time gives
\begin{equation}
\frac{\partial \mathbf{j}}{\partial t} = \frac{\dot{\rho}}{\rho}\mathbf{j} + \frac{q\rho}{m} \left[ \hbar \nabla\left(\frac{\partial}{\partial t}(\theta\coth R)\right) - q\frac{\partial \mathbf{A}}{\partial t} \right].
\label{eq:firstlondon}
\end{equation}
Using the electric field \(\mathbf{E} = -\nabla\phi_e - \partial_t \mathbf{A}\) (where \(\phi_e\) is the electric scalar potential), one may also write this as
\begin{equation}
\frac{\partial \mathbf{j}}{\partial t} = \frac{\dot{\rho}}{\rho}\mathbf{j} + \frac{q^2\rho}{m}\mathbf{E} + \frac{q\hbar\rho}{m} \nabla\left( \frac{\partial}{\partial t}(\theta \coth R) + \frac{q}{\hbar}\phi_e \right).
\label{eq:firstlondonE}
\end{equation}

This relation departs qualitatively from the usual first London equation. In addition to the ordinary electric-field response, there appears a convective term proportional to \((\dot{\rho}/\rho)\mathbf{j}\), as well as a geometric acceleration term involving \(\partial_t(\theta\coth R)\). These additional structures imply that collective amplitude dynamics can contribute directly to the electrodynamic response of the condensate.

The second London equation is modified even more strongly. Taking the curl of Eq.~\eqref{eq:supercurrent}, one finds
\begin{equation}
\nabla \times \mathbf{j} = -\frac{q^2}{m}\nabla \times (\rho \mathbf{A}) + \frac{q\hbar}{m}\nabla \times \left[\rho \nabla(\theta\coth R)\right].
\label{eq:curlj}
\end{equation}
Expanding this expression gives
\begin{equation}
\nabla \times \mathbf{j} = -\frac{q^2\rho}{m}\mathbf{B} - \frac{q^2}{m}(\nabla\rho \times \mathbf{A}) + \frac{q\hbar}{m} \left[ \nabla\rho \times \nabla(\theta\coth R) \right].
\label{eq:secondlondon}
\end{equation}
By invoking Amp\`ere's law (\(\nabla \times \mathbf{B} = \mu_0 \mathbf{j}\)) and taking the curl of both sides, we can map out the structural deformation of the magnetic screening field directly in terms of the underlying particle density \(\rho = R^2\)
\begin{equation}
    \nabla ^2 \mathbf{B}=-\mu _{0}\frac{q\hbar}{m}\nabla \times [n \nabla(\theta \coth\sqrt{\rho})]+\mu _{0}\frac{q^2}{m}(\rho\mathbf{B} - \nabla \rho\times \mathbf{A}).
\end{equation}

Equation~\eqref{eq:secondlondon} makes clear that the Meissner response is no longer controlled by a single scalar screening length. Instead, magnetic field screening becomes sensitive to the local amplitude texture and to the relative orientation of density and phase gradients. In regions where the order parameter is nearly uniform, one expects approximately conventional London behavior. In the presence of strong amplitude structure, however, the magnetic response should become spatially textured and, in general, depart from simple exponential screening due to the source terms generated by cross-gradient coupling.

\subsection{Flux Quantization and Zero-Current States}

The nontrivial phase manifold also modifies the topological sector of the theory. Since the order parameter must remain single-valued, the generalized phase must satisfy
\begin{equation}
\oint \nabla(\theta \coth R)\cdot d\mathbf{l} = 2\pi n.
\label{eq:singlevalued}
\end{equation}
This replaces the usual winding condition on the ordinary phase alone. The quantized object is therefore not \(\theta\) itself, but the composite field \(\theta\coth R\). When electromagnetic coupling is included, the relevant gauge-invariant circulation is
\begin{equation}
\oint \left[ \hbar \nabla(\theta\coth R) - q\mathbf{A} \right]\cdot d\mathbf{l} = m \oint \mathbf{v}_s \cdot d\mathbf{l},
\label{eq:gaugecirc}
\end{equation}
which yields the modified flux relation
\begin{equation}
\Phi = n\Phi_0 - \frac{m}{q} \oint \mathbf{v}_s \cdot d\mathbf{l}, \qquad \Phi_0 = \frac{h}{q}.
\label{eq:flux}
\end{equation}
For Cooper-pair condensates, one recovers the usual flux unit \(\Phi_0 = h/2e\), but the interpretation of winding is no longer conventional. Topological circulation can now be redistributed among ordinary phase accumulation, amplitude-induced geometric twisting, and electromagnetic flux. This suggests that vortex-like excitations in the present framework may differ qualitatively from those of standard Ginzburg--Landau theory. In particular, one may expect internal amplitude shells, split-core structures, or annular current textures associated with the singular geometry of the phase manifold.

A particularly interesting consequence of the generalized phase manifold is the existence of nontrivial stationary states in which the gauge-covariant current vanishes despite the presence of nonuniform amplitude and nontrivial electromagnetic structure. Setting
\begin{equation}
\mathbf{v}_s = 0
\label{eq:zerov}
\end{equation}
implies
\begin{equation}
\hbar \nabla(\theta\coth R) = q\mathbf{A}.
\label{eq:compensation}
\end{equation}
Integrating this relation yields a compensation condition of the form
\begin{equation}
\theta(\mathbf{r}) = \tanh R(\mathbf{r}) \left[ K + \frac{q}{\hbar}\oint_{\mathcal{C}} \mathbf{A}\cdot d\mathbf{l} \right],
\label{eq:magic}
\end{equation}
where \(K\) is an integration constant and \(\mathcal{C}\) is a chosen path. This relation identifies a class of states in which the internal geometric momentum exactly balances the ordinary gauge momentum. Such configurations are not currentless in a trivial sense; rather, they are currentless because the condensate self-organizes its phase manifold so as to absorb the applied gauge stress. The electromagnetic environment is therefore not merely screened by current flow, but can in principle be accommodated by geometric rearrangement of the order parameter itself.

\subsection{Gross-Pitaevskii and Klein-Gordon equations}

Up to this point we have left the standard Schr\"odinger equation unaltered. We now apply the same hydrodynamic transformation to the Gross–Pitaevskii equation \cite{PitaevskiiStringari}, which governs the dynamics of a weakly interacting Bose–Einstein condensate. This equation incorporates a mean‑field interaction term proportional to the density and therefore provides a natural extension of the previous analysis
\begin{equation}
    i\hbar \frac{\partial \Psi}{\partial t}=-\frac{\hbar^2}{2m}\nabla^2 \Psi+V\Psi+g|\Psi|^2 \Psi.
\end{equation}

By performing the squeezed Madelung substitution and separating the real and imaginary parts, we obtain the corresponding Eulerian equation for the condensate velocity field. This equation generalizes the classical Euler equation by including both the external potential and the quantum pressure term
\begin{equation}
    m\left(\frac{\partial \mathbf{v}}{\partial t}+\mathbf{v}\cdot\nabla \mathbf{v}\right)=-\nabla (Q+V+g\rho).
\end{equation}

This momentum equation can also be written in conservative form, highlighting the flux of momentum density and making the analogy with classical fluid dynamics more explicit
\begin{equation}
    \frac{\partial (\rho \mathbf {v})}{\partial t}+\nabla \cdot (\rho \mathbf {v}\otimes \mathbf {v})=-\rho \nabla (Q+V+g\rho).
\end{equation}

Now, by taking the time derivative of the continuity equation and the gradient of the momentum equation, we eliminate the current term \cite{Khalatnikov}. This procedure yields a second-order equation for the density alone, which is the natural starting point for studying linear perturbations and collective excitations
\begin{equation}
    \frac{\partial^2 \rho}{\partial t^2}=\nabla_i \nabla_j (\rho v_i v_j)+\frac{1}{m}\nabla \cdot \left[\rho\nabla (Q+V+g\rho)\right].
\end{equation}

The linearization of the theory about a uniform background would reveal whether the amplitude-phase coupling generates new collective modes or mixed electrodynamic resonances. To analyze small fluctuations, we linearize the system around a stationary background by writing \(\rho \rightarrow \rho _0+\delta \rho\) and \(\mathbf{v} \rightarrow \mathbf{v} _0+\delta \mathbf{v} \). The perturbation of the velocity field is then approximated by
\begin{equation}
   \mathbf{v} \approx \mathbf{v}_0-\frac{\hbar \theta_0}{2m\sqrt{\rho_0}\sinh^2\sqrt{\rho_0} }\delta\rho .
\end{equation}
Substituting these expressions into the linearized equations and keeping only first‑order terms, we find the following wave‑type equation for the density perturbation
\begin{equation}
      \frac{\partial^2 \delta\rho}{\partial t^2}\approx-\frac{\hbar ^2}{4m^2}\nabla^4 \delta\rho+\left(\mathbf{v}_0^2+\frac{g\rho_0}{m}\right)\nabla ^2 \delta \rho.
\end{equation}
From this it follows that the real part of the excitation frequency satisfies the dispersion relation,
\begin{equation}
\omega_R ^2=\frac{B}{2}+\sqrt{\left(\frac{B}{2}\right)^2+C^2},
\end{equation}
where the coefficient \(B\) collects the contributions from the quantum pressure and the mean‑field interaction,
\begin{equation}
    B=\frac{\hbar^2 k^4}{4m^2}+\left(\frac{g\rho_0}{m}+\mathbf{v}_0^2\right)k^2,
\end{equation}
and the coefficient \(C\) encodes the effects of spatial inhomogeneities in the external potential as well as corrections arising from the background flow
\begin{equation}
    C=\frac{\mathbf{k}\cdot\nabla V}{2m}-\frac{\hbar (\mathbf{k}\cdot\mathbf{v}_0)\theta_0 \rho_0^{1/2}k^2 }{2m\sinh^2\rho_0 ^{1/2}}.
\end{equation}

We now repeat the same hydrodynamic procedure, but this time applied to the Klein-Gordon equation viewed simply as a wave equation describing a massive particle and not the evolution of the scalar field. Although this equation has a formal structure distinct from the Schr\"odinger and Gross–Pitaevskii cases, we can still separate the wavefunction into amplitude and phase and extract density and velocity‑like quantities
\begin{equation}
   ( \Box^2 +m^2)\Psi=0.
\end{equation}
From the conserved current associated with this equation, we identify the probability density. Writing the wavefunction in the squeezed hyperbolic form, the density becomes
\begin{equation}
    \rho=R^2 \frac{\partial \phi}{\partial t}=R^2 \left(\coth R\frac{\partial \theta}{\partial t}-\frac{\theta}{\sinh^2R}\frac{\partial R}{\partial t}\right).
\end{equation}
This expression shows that the density depends not only on the time variation of the phase but also on how the amplitude changes in time. In the same way, we can define a velocity field by taking the ratio of spatial and temporal derivatives of the phase. This gives a natural analogue of the velocity field in this wave‑based description
\begin{equation}
    \mathbf{v}=-c^2\frac{\nabla\phi}{\frac{\partial \phi}{\partial t}}=-c^2 \left(\frac{\frac{1}{2}\sinh 2R\nabla\theta-\theta \nabla R}{\frac{1}{2}\sinh 2R\frac{\partial \theta}{\partial t}-\theta \frac{\partial R}{\partial t}}\right).
\end{equation}

Once the amplitude and phase have been separated, the Klein-Gordon equation yields a Hamilton-Jacobi-type relation. This equation plays the same role as before; it governs the evolution of the phase and contains an additional term involving the amplitude, which acts as a quantum correction
\begin{equation}
    \frac{1}{c^2}\left(\nabla\theta \coth R-\frac{\theta}{\sinh^2R}\nabla R\right)^2-\frac{1}{c^4}\left|\frac{\partial \theta}{\partial t} \coth R-\frac{\theta}{\sinh^2R}\frac{\partial R}{\partial t}\right|^2=m^2c^2+\Box ^2 R.
\end{equation}
This relation tracks how the squeezed phase manifold behaves under certain constraints, revealing that high-frequency time modulations of the amplitude act as active energy-shifting potentials within the moving frame of the quantum fluid.

This Hamilton–Jacobi relation demonstrates how the squeezed hyperbolic ansatz deforms the fundamental spacetime tracking of the quantum fluid. In the standard Madelung decomposition of the Klein-Gordon equation, the squared four-gradient of the phase balances the rest mass plus a d'Alembertian quantum potential ($\Box^2 R / R$). Under our hyperbolic transformation, however, the local phase gradients and temporal evolution are inextricably bound to the amplitude field $R$ and its variations across both space and time. The second term on the left-hand side reveals that high-frequency temporal modulations of the amplitude act as active energy-shifting potentials, altering the effective local rest mass of the field within its moving frame. Concurrently, the standard quantum potential is structurally absorbed into the d'Alembertian operator $\Box^2 R$ on the right-hand side. This demonstrates that the squeezed phase manifold adapts dynamically to preserve covariant transport properties, establishing a self-consistent flow where spatial density textures and temporal fluctuations directly regulate the local inertia of the quantum system.

\subsection{Further possible applications}

Several extensions of the present work appear especially natural. A fully variational treatment would begin from a generalized Ginzburg-Landau \cite{PitaevskiiStringari} free energy in which the nonlinear phase manifold is incorporated directly into the kinetic term. Such a formulation would make it possible to determine whether the singular amplitude layers survive as exact structures or are regularized into finite-width internal interfaces. To study this variationally, the system can be modeled using the effective hydrodynamic Lagrangian density \(\mathcal{L}\),
\begin{equation}
    \mathcal{L}=- \rho\hbar \frac{\partial {(\theta \coth\sqrt{\rho}})}{\partial t}-\frac{\rho}{2m}\left[\hbar \nabla (\theta \coth \sqrt{n})-q\mathbf{A}\right]^2-\frac{\hbar ^2}{8mn}(\nabla \rho)^2-V\rho.
\end{equation}
This formulation provides a direct framework for deriving the coupled Euler-Lagrange equations for both the auxiliary field \(\theta\) and the physical density.

 A second important direction would be the explicit construction of static vortex solutions, which would clarify the topological sector of the theory and test whether annular, shell-like, or geometrically compensated flux tubes are energetically preferred. Finally, a nonlocal extension in the spirit of Pippard electrodynamics \cite{Pippard1953} may provide a natural mechanism for smoothing the singular walls while preserving the essential amplitude-dependent geometry. Under a generalized kernel \(K(|\mathbf{r}-\mathbf{r}'|)\), the localized supercurrent constitutive law expands to
\begin{equation}
 \mathbf{j}( \mathbf{r} )=-\int d^3 \mathbf{r}' K(|\mathbf{r}-\mathbf{r}'|) \left[\mathbf{A}(\mathbf{r}')-\frac{\hbar}{q}\nabla (\theta \coth \sqrt{\rho})_{\mathbf{r}'}\right].
\end{equation}
This nonlocal modification effectively averages out the short-wavelength singularities near sharp amplitude boundaries, rendering the geometric phase transitions physically smooth. In this non-local formulation, $K(|\mathbf{r}-\mathbf{r}'|)$ represents an isotropic scalar kernel that acts as a spatial weight, generalizing the constitutive relation in the spirit of Pippard's electrodynamics. Rather than assuming a strictly local response where the supercurrent depends entirely on fields at the singular point $\mathbf{r}$, the kernel accounts for a characteristic coherence length—analogous to the Pippard coherence length $\xi_0$ in conventional superconductors—over which the phase-entangled condensate can sample the electromagnetic and geometric environment. Because the kernel depends solely on the isotropic Euclidean distance $|\mathbf{r}-\mathbf{r}'|$, it preserves the fundamental rotational invariance of the bulk medium. Mathematically, this integral convolution operates as a spatial low-pass filter; it effectively averages out and regularizes the short-wavelength singularities that naturally arise near sharp amplitude boundaries where $R \rightarrow 0$ or where $\coth R$ diverges, rendering the macroscopic geometric phase transitions physically smooth across spatial interfaces.

Beyond macroscopic order parameters, the $\coth$-Madelung framework offers compelling avenues for resolving deep conceptual difficulties in quantum tunneling and early-universe cosmology. One natural direction concerns quantum tunneling and related phenomena such as the Hartman effect; in quantum barrier penetration, the traditional phase-gradient velocity field drops to zero or becomes purely evanescent, giving rise to the controversial Hartmann effect where tunneling times appear independent of barrier width, implying unphysical or instantaneous propagation. Because the coth-Madelung transformation incorporates an explicit amplitude-gradient component ($\nabla R$), it establishes a non-vanishing, amplitude-sensitive velocity field directly inside classically forbidden regions. A second area of interest is quantum cosmology. Because Bohmian and Madelung-type decompositions play an important role in the semiclassical analysis of the Wheeler-DeWitt equation, the \(\\coth\)-Madelung representation may offer a novel way of encoding amplitude-dependent geometric effects in minisuperspace or superspace and a fresh perspective for the emergence of quantum cosmological behavior. 

\section{Discussion and Outlook}

The $\coth$-Madelung transform introduced here suggests a broader perspective on quantum hydrodynamics and superconducting order. In the conventional picture, the amplitude of the wavefunction is often treated as a relatively passive scalar field whose primary role is to determine density and modulate stiffness. By contrast, the present construction shows that once the phase is allowed to depend nontrivially on the amplitude, the density field becomes a genuinely geometric degree of freedom. It does not merely weigh the current; it shapes the manifold along which current is permitted to flow.

This geometric viewpoint leads naturally to a number of physical expectations. The first is that the Meissner effect should become spatially textured, particularly in strongly inhomogeneous or frustrated superconductors. The second is that amplitude collective modes should acquire direct electrodynamic relevance, potentially coupling to current response more strongly than in conventional superconductors. The third is that topological defects should inherit internal amplitude-phase structure, leading to vortex morphologies that differ from the usual paradigms. Finally, the existence of geometric compensation states suggests that certain superconducting configurations may be stabilized not by uniform phase rigidity, but by adaptive rearrangement of the order-parameter manifold itself.

At a broader level, the framework developed here is perhaps best viewed not as a replacement for conventional superconducting hydrodynamics, but as a candidate effective theory for regimes in which phase and amplitude cannot be cleanly disentangled. This may be particularly relevant in systems where superconductivity coexists with strong spatial modulation, competing orders, or anomalous phase stiffness. In such settings, the order parameter may not live on a simple rigid phase circle, but on a curved and dynamically structured manifold whose geometry materially influences transport and topology.

\section{Conclusion}

We have shown that the nonlinear substitution \(\Psi = R e^{\imath\theta \coth R}\) defines a generalized hydrodynamic representation in which amplitude and phase are intrinsically coupled. The resulting velocity field contains an explicit density-gradient contribution, implying that superflow is governed not solely by phase winding but by the geometry of the amplitude landscape. When the wavefunction is interpreted as a superconducting order parameter, this coupling leads to generalized London equations in which the electromagnetic response is spatially textured and sensitive to internal density structure.

The principal conceptual outcome of the construction is that superconducting rigidity need not be understood as a purely phase-driven phenomenon. Instead, the condensate may behave as a geometrically adaptive medium whose screening, current response, and topological organization are all shaped by the amplitude-dependent phase manifold on which it resides. In that sense, the present framework points toward a more general hydrodynamic principle, namely that the electrodynamics of a quantum condensate may depend not only on the magnitude of its order, but on the internal geometry through which that order is allowed to propagate. The resulting hydrodynamic equations define a class of quantum fluids whose inertia and transport properties are governed by a nontrivial phase manifold over density space. Flux quantization is likewise altered, with topological winding distributed across both phase and amplitude geometry. We further identify a force-free compensation condition, \(\theta \propto \tanh R\), under which the internal geometric momentum can cancel the conventional phase current and, in gauge-covariant form, offset electromagnetic contributions as well, ensuring that superflow is shaped dynamically by local amplitude textures. Whether the resulting framework ultimately describes an effective phenomenology, a deeper hydrodynamic layer of quantum theory, or merely a mathematically suggestive extension remains an open question. Nevertheless, the appearance of hyperbolic statistical kernels, amplitude-sensitive transport, and generalized osmotic dynamics indicates that such deformations may deserve further investigation within the broader foundations of quantum hydrodynamics.

\end{document}